\documentclass[showpacs,preprintnumbers,amsmath,amssymb]{revtex4}
\usepackage[dvips]{graphicx,color}
\usepackage{epsf}
\usepackage{color}
\newcommand{\be}{\begin{equation}}
\newcommand{\ee}{\end{equation}}

\newcommand{\tr}{{\rm Tr\,}}

\newcommand{\rmd}{{\rm d}}
\begin{document}

\title{Scalar-graviton interaction in the noncommutative space}

\author{F. T. Brandt and M. R. Elias-Filho}

\affiliation{Instituto de F\'{\i}sica, Universidade de S\~ao Paulo,
S\~ao Paulo, SP 05315-970, BRAZIL}

\date{\today \\ }

\begin{abstract}
We obtain the leading order interaction between the graviton and
the neutral scalar boson in the context of noncommutative field theory.
Our approach makes use of the Ward identity associated with the 
invariance under a subgroup of symplectic diffeomorphisms.
\end{abstract}


\maketitle

The formulation of field theories in noncommutative spaces 
has been investigated from various point of 
view \cite{Szabo:2001kg,Douglas:2001ba}. The
standard approach which leads to the replacement the usual product of fields by 
the Gr\"onewold-Moyal product \cite{Groenewold:1946,Moyal:1949sk} 
has been employed in the case of scalar field theories as well as 
in gauge theories with symmetry group U(N). In the later case, it is 
possible to obtain a gauge invariant $\star$-deformed formulation
which has the gauge symmetries similarly to the commutative case.
In this brief report we will present a simple derivation of
a $\star$-deformed formulation of the interaction vertices of
gravitons and massless neutral bosons. This is the simplest model
involving both gravity and noncommutativity.

It is well known that, in the commutative space, the interaction vertices 
of gravitons with other quanta (including the self-interactions of gravitons) can be
obtained from the weak field expansion of the action formulated in 
a curved space. The action is invariant under gauge transformations
which in turn implies the existence of Ward identities between
the vertices. Conversely, the Ward identities
can be employed in order to determine all the vertices starting from
the free quadratic term. This approach would
be the only alternative in cases when one does not know
the closed form of the action.
Although this is hardly the case in the commutative space,
this approach may be an (instructive) alternative to derive the interaction vertices
between gravitons and massless neutral bosons in the noncommutative space.

In order to put this idea into practice, we would have to know, 
first of all, what is the gauge symmetry which relates 
the interactions vertices. It is not
obvious what would be the symmetry since the
invariance under coordinate transformations does not seem to be compatible
with the commutation relation
\be\label{eq1}
\left[x^\mu,x^\nu\right]_\star = x^\mu\star x^\nu  -  x^\nu\star x^\mu 
=i \theta^{\mu\nu},
\ee
where $\theta^{\mu\nu}$ is independent of coordinate, and the 
$\star$-product is defined as
\begin{eqnarray}\label{moyal1}
f(x)\star g(x) = f(x)~\exp\left(\frac i2\,\overleftarrow{\partial_\mu}\,
\theta^{\mu\nu}\,\overrightarrow{\partial_\nu}\right)~g(x) .
\end{eqnarray}
However, it has been recently pointed out \cite{Calmet:2005qm} that 
a restricted class of coordinate transformations of the form
\begin{subequations}\label{transcoord}
\be
x^\mu\rightarrow x^{\prime\mu} = x^\mu - \tilde \omega^\mu(x)
\ee
with infinitesimal parameter
\be\label{otil}
\tilde\omega^\mu(x) = -i\,[x^\mu,\Lambda(x)]_\star = 
\theta^{\mu\alpha}\,\partial_\alpha \Lambda,
\ee
\end{subequations}
maintains the commutation relation \eqref{eq1} unchanged.
This can be easily verified by directly computing the transformed commutator
$\left[x^{\prime \mu},x^{\prime \nu}\right]_\star$ with the help of Eqs. \eqref{transcoord} 
and taking into account the Jacobi identity
\be
[x^\mu,[\Lambda,x^\nu]]+[\Lambda,[x^\nu,x^\mu]]+[x^\nu,[x^\mu,\Lambda]] = 0
\ee
as well as $[\Lambda,[x^\nu,x^\mu]] = 0$.

Let us now write down the most general form that action would have at
any given order. It can be expressed as the sum of the following
perturbative terms
\begin{subequations}\label{acao_S(i)}
\begin{eqnarray}
\label{acao0}
S^{(0)} & = & \frac{1}{2} \int\rmd^4 x\, 
\eta^{\mu\nu}\,\partial_\mu\phi\,\partial_\nu \phi ,\\
S^{(1)} & = & \frac{\kappa}{2} \int\rmd^4 x\, 
C^{(1)\mu\nu\mu_1\nu_1}\,
h_{\mu_1\nu_1}\star\partial_\mu\phi\star \partial_\nu \phi,\\
  \vdots \nonumber \\  
S^{(n)} & = &\frac{\kappa^n}{2} \int\rmd^4 x\, \label{5c}
C^{(n)\mu\nu\mu_1\nu_1 \dots \mu_n\nu_n}\,
h_{\mu_1\nu_1}\star \cdots \star h_{\mu_n\nu_n}\star 
\partial_\mu\phi\star \partial_\nu \phi + \cdots,
\end{eqnarray}
\end{subequations}
where $\kappa$ is the coupling constant which arises in the usual weak
field expansion $g_{\mu\nu} = \eta_{\mu\nu} + \kappa h_{\mu\nu}$ and
the graviton and scalar fields are $h_{\mu\nu}(x)$ and $\phi(x)$ respectively. 
The ellipses in Eq. \eqref{5c} denote terms involving permutations of
the gravitational fields and gradients which are not cyclically equivalent.
From dimensional analysis the structures $C^{(n)\mu\nu\mu_1\nu_1 \dots \mu_n\nu_n}$
have to be dimensionless. The simplest possibility is to assume that
$C^{(n)\mu\nu\mu_1\nu_1 \dots \mu_n\nu_n}$ is a combination of
products of the Minkowski metric. For instance,
\be\label{C1}
C^{(1)}= a\, \eta^{\mu\nu}\eta^{\mu_1\nu_1 }+b \,\big(\eta^{\mu\mu_1}\eta^{\nu\nu_1}
+\eta^{\mu\nu_1}\eta^{\nu\mu_1}\big),
\ee
where we have taken into account the symmetry $h_{\mu_1\nu_1} = h_{\nu_1\mu_1}$. 
This in turn implies that the first order term can be written as
\be
\label{acao1}
S^{(1)} =  \frac{\kappa}{4} \int\rmd^4 x\, 
C^{(1)\mu\nu\mu_1\nu_1}\, h_{\mu_1\nu_1} \star 
\left\{\partial_\mu\phi,\partial_\nu \phi\right\}_\star,
\ee
where we have introduced the Moyal anti-commutator 
$\left\{A,B\right\}_\star=A\star B + B\star A$.
It remains to determine the numerical values of $a$ and $b$ in Eq. \eqref{C1}.

The restricted class of transformations \eqref{transcoord} implies that
\be
\det\left(\frac{\partial x^{\prime\mu}}{\partial x^{\nu}}\right)
=\exp{\left[\tr\log(\delta^\mu_\nu-\partial_\nu\tilde\omega^\mu)\right]}
=\exp{(-\partial_\mu\tilde\omega^\mu)}=1,
\ee
so that we can assume, consistently with the invariance of $\sqrt{-g}$, the weak
field condition $h^\mu_\mu=0$, which in the present context 
is a noncommutative version of unimodular gravity \cite{Einstein:1919gv}. 
Therefore, the first term in Eq. \eqref{C1} will not contribute and
the only parameter to be determined is $b$  (in the commutative 
theory both terms would contribute and they can be determined, 
from the Ward identity associated with the coordinate transformation).
Therefore, the first order action can be reduced to the form
\be\label{act1}
S\approx \frac{1}{2} \int\rmd^4 x\, \left(
\eta_{\mu\nu}\partial^\mu\phi\star\partial^\nu\phi + {b}\,{\kappa}
h_{\mu\nu}\left\{\partial^\mu\phi,\partial^\nu\phi\right\}_\star \right),
\ee
where value of $b$ can now be determined from the gauge invariance of $S$. 

The restricted coordinate transformations yields the following 
gauge transformations
\begin{subequations}\label{gaugetrans}
\begin{eqnarray}
\label{gauge1a}
\kappa \, \delta h_{\mu\nu} & = & \partial_\mu \tilde \omega_\nu + 
                       \partial_\nu \tilde \omega_\mu 
+{\cal O}(\kappa)
=\theta_{\mu\lambda}\partial^\lambda\partial_\nu\Lambda +
 \theta_{\nu\lambda}\partial^\lambda\partial_\mu\Lambda
+{\cal O}(\kappa),
\\  \label{gauge1b}
\delta\phi & = & \frac{1}{2}\left\{\tilde \omega^\alpha,\partial_\alpha \phi\right\}_\star=
\frac 1 2 \theta^{\alpha\beta}
\left\{\partial_\beta\Lambda,\partial_\alpha\phi\right\}_\star .
\end{eqnarray}
\end{subequations}
We have chosen to write down Eq. \eqref{gauge1b} in terms of the anti-commutator,
to assure that the gauge transformations have a well defined behavior under
$\theta^{\mu\nu}\rightarrow - \theta^{\mu\nu}$. Imposing the condition
$\delta S = 0$ we obtain from Eqs. \eqref{act1} and \eqref{gaugetrans}
\be\label{dS}
\frac 1 2 \theta^{\alpha\beta}  \eta^{\mu\nu}\left(
\int\rmd^4 x \,
\partial_\nu\phi\star \partial_\mu
\left\{\partial_\beta\Lambda,\partial_\alpha\phi\right\}_\star  
+ 2 b\, 
\int\rmd^4 x \,
\partial_\mu \partial_\beta \Lambda 
\star  \left\{\partial_\alpha \phi , 
\partial_\nu \phi\right\}_\star\right) = 0.
\ee
Expanding the first term we obtain
\be
\frac 1 2 \theta^{\alpha\beta}\eta^{\mu\nu}\int\rmd^4 x \left[
\partial_\nu \phi\star
\partial_\mu\partial_\beta\Lambda\star\partial_\alpha\phi +
\partial_\nu \phi\star
\partial_\beta\Lambda\star\partial_\mu\partial_\alpha\phi +
\partial_\nu \phi\star
\partial_\mu\partial_\alpha\phi\star\partial_\beta\Lambda +
\partial_\nu \phi\star
\partial_\alpha\phi\star\partial_\mu\partial_\beta\Lambda 
\right].
\ee
Performing integration by parts in the second term and using the
antisymmetry of $\theta^{\alpha\beta}$, yields
\be
\frac 1 2 \theta^{\alpha\beta}\eta^{\mu\nu} \int\rmd^4 x \left[
\partial_\nu \phi\star
\partial_\mu\partial_\beta\Lambda\star\partial_\alpha\phi -
\partial_\alpha\partial_\nu \phi\star
\partial_\beta\Lambda\star\partial_\mu\phi +
\partial_\nu \phi\star
\partial_\mu\partial_\alpha\phi\star\partial_\beta\Lambda +
\partial_\nu \phi\star
\partial_\alpha\phi\star\partial_\mu\partial_\beta\Lambda 
\right].
\ee
Using the cyclicity property of the $\star$-product the second and the
third terms in the above equation cancel each other and we are left
with
\be
\frac 1 2 \theta^{\alpha\beta}\eta^{\mu\nu} \int\rmd^4 x \left[
\partial_\mu\partial_\beta\Lambda\star
\partial_\alpha\phi\star\partial_\nu \phi +
\partial_\mu\partial_\beta\Lambda \star 
\partial_\nu \phi\star
\partial_\alpha\phi
\right]
=\frac 1 2 \theta^{\alpha\beta}\eta^{\mu\nu} \int\rmd^4 x 
\partial_\mu\partial_\beta\Lambda\star
\left\{\partial_\alpha\phi, \partial_\nu \phi\right\}_\star ,
\ee
where we have used again the  cyclicity property of the $\star$-product.
Therefore the two terms in Eq. \eqref{dS} have the same structure.
Replacing the first term of  Eq. \eqref{dS} by the previous
expression, yields the relation
\be\label{15}
\left(\frac 1 2  +  b\right)
\theta^{\alpha\beta}  \eta^{\mu\nu}
\int\rmd^4 x \,      
\partial_\mu \partial_\beta \Lambda 
\star  \left\{\partial_\alpha \phi , 
\partial_\nu \phi\right\}_\star = 0.
\ee
This relation shows that the value of $b$ which makes the action gauge
invariant is $b=-1/2$.
A similar calculation shows that the value of the 
parameter $a$ in \eqref{C1} is undetermined. Technically this happens
because, instead of a relation like \eqref{15}, the equation for $a$
would be multiplied by a vanishing expression. Thus, the condition
$h_\mu^\mu=0$ is important in order to fix the unique form of the action.

This result shows that the restricted gauge invariance can give
information on the form of the interactions. With additional conditions
such as the most simple form assumed for the structures 
$C^{(n)\mu\nu\mu_1\nu_1 \dots \mu_n\nu_n}$ we were able to find the
leading order contribution to the action. Of course one could in
principle consider more complicated scenarios with the introduction of
dimensionfull parameters as well as more complicated structures
involving derivatives and $\theta^{\mu\nu}$. 
However, our objective here was to exhibit the simplest scenario which
is compatible with the restricted gauge invariance.

Our approach may be considered complementary to the ones which
employs power expansions in $\theta^{\mu\nu}$. It is interesting
to note that the result in Eq. \eqref{act1} involves 
only the anticommutator $\left\{\partial_\mu\phi,\partial_\nu\phi\right\}$
so that the expansion in powers of $\theta^{\mu\nu}$ does not contain
a term linear in $\theta^{\mu\nu}$. This property has also been
verified in the context of pure gravity \cite{Mukherjee:2006nd}.

When the full effects of gravity are considered, so that the weak
field expansion cannot be used, one would have to generalize the $\star$-product
in such a way that the coordinate dependence of $\theta^{\mu\nu}$ has to be taken
into account in a generally covariant fashion. For instance, the
covariant derivative of $\theta^{\mu\nu}$ is taken to be zero.
This has been recently investigated in \cite{Harikumar:2006xf} where 
the proposed from of the scalar-graviton action has 
a power expansion in $\theta^{\mu\nu}$ and $h^{\mu\nu}$ which coincides
with our leading order result in the special case of constant $\theta^{\mu\nu}$.
Thus, our weak field result, which seems to be compatible with the usual
$\star$-product deformation of unimodular gravity, is also consistent with 
the more general formulations. Our approach may be suitable in the
physical situations which can be described in terms 
of the exchange of single gravitons in a flat background.

Finally, we remark that although the individual actions
$S^{(n)}$ are not globally Lorentz invariant we still have a residual local
symmetry which suffice to shape the form of the interaction vertices.
This is analogous to what happens for example in the case of the
effective action in thermal gravity where the Lorentz symmetry is
broken by the thermal bath but the effective vertices are all related
by Ward identities associated with the local gauge symmetry in the
high temperature limit \cite{Brandt:1992qn}.

\bigskip

\noindent{\bf Acknowledgment}

The authors would like to thank CAPES, CNPq, and FAPESP, Brazil,
for financial support.


\end{document}